\documentclass[12pt]{article}
\usepackage[utf8]{inputenc}
\usepackage{amsmath,setspace,geometry}
\usepackage{amsthm}
\usepackage{amsfonts}
\usepackage{amssymb}
\usepackage[shortlabels]{enumitem}
\usepackage{rotating}
\usepackage{pdflscape}
\usepackage{graphicx}
\usepackage{bbm}
\usepackage[dvipsnames]{xcolor}
\usepackage{hyperref}
\hypersetup{colorlinks=true, linkcolor= BrickRed, citecolor = BrickRed, filecolor = BrickRed, urlcolor = BrickRed, hypertexnames = true}
\usepackage[]{natbib} 
\bibpunct[:]{(}{)}{,}{a}{}{,}
\usepackage[english]{babel}
\usepackage{float}
\usepackage{caption}
\usepackage{subcaption}
\usepackage{booktabs}
\usepackage{pdfpages}
\usepackage{threeparttable}
\usepackage{lscape}
\usepackage{bm}
\usepackage{multirow}
\usepackage{booktabs}
\usepackage{adjustbox}
\bibpunct[:]{(}{)}{,}{a}{}{,}
\newtheorem{proposition}{Proposition}
\newtheorem{corollary}{Corollary}
\theoremstyle{definition}
\newtheorem{remark}{Remark}
\usepackage{setspace}
\title{The Value of a Chance: Task Concentration and Talent Discovery in Team Production
\thanks{I am grateful to Ryohei Hayashi, Ryo Nakajima, and Fumio Ohtake for their valuable comments and suggestions that greatly improved this paper. I thank Reio Tanji, a baseball analyst, for sharing his domain expertise and providing insightful feedback from a practitioner's perspective. I also gratefully acknowledge comments from participants at the Kansai Labor Workshop (March 2025) and the 19th Annual Conference of the Association of Behavioral Economics and Finance (December 2025). This work was supported by JST SPRING, Grant Number JPMJSP2123.
}}
\author{Masaya Nishihata
\thanks{Graduate School of Economics, Keio University, 2-15-45 Mita, Minato-ku, Tokyo 108-8345, Japan (e-mail: \href{mailto:}{nishihata-masaya@keio.jp}).}
}
\date{
\today
}

\begin{document}

\maketitle

\begin{abstract}
    Organizations often concentrate scarce, high-value tasks on proven performers, but doing so may limit opportunities to develop and learn about alternative workers. We study this trade-off using temporary injuries to high-performing Major League Baseball players, which generate plausibly exogenous reallocations of playing opportunities. Tracking allocation and performance before, during, and after these absences, we find clear evidence of crowd-out: when a high performer becomes unavailable, same-team, same-position coworkers receive substantially more opportunities. Yet we find little evidence that the additional experience raises batting productivity during the absence or after the high performer returns, suggesting that learning by doing is not the main dynamic benefit. Instead, performance surprises during the absence predict subsequent employment and playing opportunities, consistent with teams using newly revealed information in later allocation decisions. Counterfactual exercises further suggest that the information generated by temporary reallocations can have positive future production value when it changes subsequent allocation, partially offsetting short-run production losses in some cases. The results highlight a general organizational trade-off: concentrating important tasks on known high performers can protect current output while slowing the discovery of capable alternatives.

\end{abstract}

\clearpage

\setstretch{1.5}

\section{Introduction} \label{sec:introduction}

Organizations routinely face the problem of allocating scarce and important tasks among workers whose abilities differ and are not always fully known. It is therefore natural to assign these tasks to workers who are already known to be highly capable. In surgery, law, or consulting, only a small number of individuals can take leading roles in important cases, procedures, or projects. Such responsibilities naturally tend to fall to proven performers. Yet task assignments do more than generate current output. They can also create two distinct forms of future value. First, workers may become more productive through learning by doing, so experience with important tasks can build human capital. Second, assigning a task to a worker generates information about that worker's ability. Giving a less-tested worker an opportunity can reveal that the worker is more, or less, capable than previously believed. Concentrating scarce opportunities on known high performers may therefore protect current production while reducing opportunities to develop and learn about alternative workers. The organizational question is not only who should perform an important task today, but also how today's allocation affects the capabilities the organization develops and the information it has available tomorrow.

High-performing coworkers are often thought to benefit their colleagues through knowledge spillovers or peer pressure \citep{mas2009peers,bandiera2010social,cornelissen2017peer}. Yet when productive opportunities are scarce, they may also limit others' access to important tasks. \cite{chalioti2026peer} formalize and document this mechanism: when task allocation is endogenous, stronger coworkers reduce junior workers' active time. In parallel, the employer-learning literature emphasizes that task exposure can generate valuable information about uncertain worker ability \citep{tervio2009superstars,pallais2014inefficient}, and \cite{hoey2026one} shows that unexpected early-career exposure improves later labor-market outcomes primarily through talent revelation. What remains less understood is how these forces interact within an organization's dynamic task-allocation problem. When opportunities are crowded out by known high performers, are the main foregone benefits developmental, informational, or both? And when a disruption generates new information about alternative workers, do organizations use it in subsequent allocation, and is its value large enough to offset the production cost of experimentation?

We study these questions using Major League Baseball (MLB), where temporary injuries to highly productive players force teams to reallocate playing opportunities. We use these disruptions to trace allocation and performance before, during, and after the absence. Following the logic of \cite{azoulay2010superstar}, we exploit injury-induced variation in the availability of high-performing players, which provides plausibly exogenous variation in teammates' playing opportunities. The temporary nature of these absences is particularly useful because many high performers return within the same season, allowing us to examine whether information generated during the absence affects subsequent allocation. MLB also provides unusually rich measures of this process: players can be linked to pre-existing same-position coworkers, playing opportunities are directly observed, performance can be separated into total production and rate-based productivity, and team-position production captures broader replacement responses including substitutions and roster adjustments. Together, these features allow us to study both the immediate crowd-out of opportunities and its dynamic consequences for information and production.

Our results reveal a sequence of responses consistent with this organizational trade-off. First, when a high-performing player becomes unavailable, playing opportunities increase sharply for the same-team, same-position coworkers defined prior to the absence, providing clear evidence of crowd-out. At the team-position level, production declines slightly during the absence and tends to be somewhat higher after the player returns. This suggestive post-return improvement raises the question of what teams may have gained during the absence. One possibility is human-capital accumulation among the coworkers who received additional opportunities. Yet although these coworkers produce substantially more while the high performer is away, we find little evidence that their productivity per unit of opportunity improves either during the absence or after the return, offering little support for learning by doing as the main mechanism. A different pattern emerges for information. Performance surprises during the absence predict subsequent allocation: workers who outperform the level expected from their pre-absence performance are more likely to remain employed and receive more starts and playing opportunities afterward. Teams therefore appear to use the information generated by temporary reallocations when making later allocation decisions. Finally, counterfactual exercises compare the production consequences of the absence with the gains from reallocating future opportunities using the newly revealed information. The production effects are highly heterogeneous, and the information revealed during the absence provides an offsetting source of future production value when it changes the preferred allocation. Taken together, the results suggest that the value of a productive opportunity lies not only in the output generated while the task is performed, but also in the information the organization obtains about who else can perform it well.

This paper contributes to three strands of literature. First, it relates to the literature on peer effects. Existing work documents both positive effects through peer pressure and knowledge spillovers \citep{falk2006clean,mas2009peers,azoulay2010superstar,yamane2015peer,frakes2021knowledge,nishihata2022heterogeneous} and negative effects through competition, motivation, or relative standing \citep{brown2011quitters,elsner2017big,li2020boon,murphy2020top,denning2023class}.\footnote{Other studies find little or no evidence of peer effects in high-skill professional settings \citep{guryan2009peer} or among university scientists \citep{waldinger2012peer}.} \cite{chalioti2026peer} identify a distinct negative externality arising from endogenous task allocation: stronger coworkers can crowd others out of scarce productive opportunities. We extend this allocation channel by studying its dynamic consequences for worker development, information production, subsequent allocation, and team production.

Second, we contribute to the literature on employer learning and talent discovery. A long-standing literature studies how uncertainty about worker productivity is resolved through observation \citep{jovanovic1979job,farber1996learning,altonji2001employer,lange2007speed,kahn2014employer}, while related work emphasizes that producing such information can itself be costly \citep{tervio2009superstars,pallais2014inefficient}. \cite{hoey2026one} shows that unexpected early-career exposure can improve later labor-market outcomes by revealing worker ability. We add an organizational allocation dimension: task assignment determines which workers generate informative signals, and we examine whether the resulting information changes subsequent allocation and is valuable enough to justify its production cost.

Third, we contribute to the literature on team production. Existing work studies how worker heterogeneity, effort, complementarities, team-specific capital, and match quality shape team outcomes \citep{hamilton2003team,gould2009interactions,arcidiacono2017productivity,jaravel2018team,cohen2024effort,nishihata2025team}. We add a dynamic information margin by showing that task allocation affects not only current production but also what the organization learns about its members. Our counterfactual analysis compares the short-run production cost of trying alternative workers with the value of the information generated by doing so.

More broadly, our findings highlight an informational value of task opportunities that extends beyond professional sports. When important tasks are scarce, worker ability is imperfectly known, and task performance is informative about ability, concentrating assignments on proven performers may limit what organizations learn about alternative workers. Similar trade-offs can arise in medicine, law, consulting, and other settings where only a few workers can take leading roles in important tasks. Our findings do not imply that organizations should routinely shift tasks away from their best performers. Rather, evaluating task allocation solely by current output can overlook the value of experimentation: giving less-proven workers meaningful opportunities may generate information that improves future allocation. The short-run cost of such experimentation should therefore be weighed against its informational value.

The remainder of the paper is organized as follows. Section~\ref{sec:conceptual} presents the conceptual framework, and Section~\ref{sec:data} describes the data. Sections~\ref{sec:empirical} and~\ref{sec:result} present the empirical strategy and main results, respectively. Section~\ref{sec:counterfactual} presents the counterfactual analysis, and Section~\ref{sec:conclusion} concludes.

\section{Conceptual framework} \label{sec:conceptual}

In this section, we develop a simple two-period model of how an organization allocates scarce productive opportunities between a worker whose productivity is known to be high and other workers. Concentrating opportunities on the high-productivity worker may raise current output, but necessarily crowds out opportunities for other workers to accumulate human capital through learning by doing and to generate information useful for future task allocation.

\subsection{Environment and Task Allocation}
\label{subsec:environment}

There are two periods, $t=1,2$, one high-productivity worker $S$, and $N$ other workers indexed by $i=1,\ldots,N$. In each period, there are $K\geq1$ related task slots. We assume $N\geq K$. We normalize the total opportunity available in each slot over a period to one. Let $x_{jt}\in[0,1]$ denote the opportunity allocated to worker $j$ in period $t$. Period-1 allocation satisfies
\[
    x_{S1}+\sum_{i=1}^{N}x_{i1}=K.
\]
The bound $x_{jt}\leq1$ implies that a worker can receive at most one unit of total task opportunity in a period. This capacity-constrained allocation setup is closely related to \cite{chalioti2026peer}, who model a principal's allocation of active time among team members under a fixed time constraint.

For each worker $i$, the organization does not perfectly observe skill $s_{i1}$ and holds a prior distribution over it with mean $a_{i1}$. For simplicity, we take the skill of $S$ to be known and assume $s_{S1}=a_{S1}>a_{i1}$ for all $i$ in period 1. This ranking need not hold in period 2.

We use the prior mean $a_{i1}$ as the organization's skill assessment. Let $f(a,x)$ denote expected current output given skill assessment $a$ and opportunity $x$. We assume $f_a>0$, $f_x>0$, $f_{ax}\geq0$, and $f_{xx}\leq0$, so that expected output increases with assessed skill and opportunity, a higher skill assessment weakly raises the marginal return to opportunity, and opportunity exhibits diminishing marginal returns.

\subsection{Human Capital and Information Production}
\label{subsec:dynamic_value}

Productive opportunities allocated to workers in period 1 can affect future organizational value through two channels. First, task experience can increase worker skill. Following \cite{kuruscu2006training}, which extends the standard human capital investment model of \cite{ben1967production}, we write worker $i$'s period-2 skill as
\[
    s_{i2}=(1-\delta)s_{i1}+g_i(x_{i1}),
\]
where $0\leq\delta<1$ and $g_i'(x)\geq0$. Thus, greater productive opportunities may raise future productivity through learning by doing or other forms of on-the-job learning.\footnote{For simplicity, we abstract from human-capital accumulation by $S$ and focus on the dynamic value of opportunities allocated to alternative workers. We also abstract from knowledge spillovers from $S$. If task allocation changes workers' exposure to $S$, the empirical productivity response captures the combined effect of own task experience and changes in knowledge spillovers.}

Second, task opportunities generate information. This channel follows the employer-learning and talent-discovery literature \citep[e.g.,][]{jovanovic1979job, tervio2009superstars, hoey2026one} in which observed performance reveals initially uncertain worker ability. Let $\mathcal{E}_{i}(x_{i1})$ denote the information structure generated by worker $i$'s period-1 performance. We assume that greater opportunity generates weakly more informative signals in the Blackwell sense \citep{blackwell1951comparison}.\footnote{Under the Blackwell order, the less informative information structure can be obtained by garbling the more informative one.} Let $y_i$ denote worker $i$'s observed period-1 performance. We normalize the performance signal so that $E[y_i\mid s_{i1},x_{i1}]=s_{i1}$. Hence, under the organization's prior, expected performance is $a_{i1}$. We refer to $y_i-a_{i1}$ as performance surprise. Observed performance updates the organization's skill assessment and may therefore affect period-2 task allocation.

Under the model's timing, period-1 performance is informative about period-1 skill, whereas human capital accumulated through period-1 task experience affects skill in period 2. Human-capital accumulation thus changes worker skill, whereas information changes the organization's assessment of skill. We allow the two channels to interact in their contribution to future organizational value.

In period 2, $S$ is unavailable with probability $q$. This parameter captures the possibility that the organization cannot rely on the high-productivity worker indefinitely, because of turnover, poaching, future absence, or other sources of worker unavailability. The baseline trade-off does not require such future unavailability; $q$ instead captures an additional reason why developing and learning about alternative workers may be valuable. Let $V^A(\mathbf{x}_1)$ and $V^U(\mathbf{x}_1)$ denote optimal expected period-2 output when $S$ is available and unavailable, respectively, where $\mathbf{x}_1=(x_{11},\ldots,x_{N1})$. These continuation values incorporate skill accumulation, information generated by period-1 performance, and optimal period-2 task allocation. Expected continuation value is
\[
    V(\mathbf{x}_1;q)
    =
    (1-q)V^A(\mathbf{x}_1)
    +
    qV^U(\mathbf{x}_1).
\]
In the unavailable state, the $K$ task slots are allocated among the other workers.

\subsection{Optimal Allocation}
\label{subsec:optimal_allocation}

The organization chooses period-1 task allocation to maximize current expected output and discounted continuation value:
\begin{equation}
\begin{split}
    \max_{x_{S1},\mathbf{x}_1}\quad&
    f(a_{S1},x_{S1})
    +\sum_{i=1}^{N}f(a_{i1},x_{i1})
    +\beta V(\mathbf{x}_1;q)\\
    \text{s.t.}\quad&
    x_{S1}+\sum_{i=1}^{N}x_{i1}=K,
    \qquad 0\leq x_{j1}\leq1,
\end{split}
\label{eq:period1_problem}
\end{equation}
where $\beta\in[0,1]$ is the weight on future output. Where differentiable, let $D_i^r(\mathbf{x}_1)\equiv \partial V^r(\mathbf{x}_1)/\partial x_{i1}$ for $r\in \{A,U\}$ denote worker $i$'s state-specific marginal continuation value. The ex ante marginal continuation value is $D_i(\mathbf{x}_1;q)\equiv (1-q)D_i^A(\mathbf{x}_1)+qD_i^U(\mathbf{x}_1)=\partial V(\mathbf{x}_1;q)/\partial x_{i1}$.

\begin{proposition}[Dynamic value and optimal allocation]
    Given $g_i'(x)\geq0$, $f_a>0$, and the Blackwell monotonicity condition stated above, the following results hold wherever $V^A$ and $V^U$ are differentiable.

    \textnormal{(i)} Greater period-1 opportunity for worker $i$ has nonnegative marginal continuation value in either availability state: $D_i^A(\mathbf{x}_1)\geq0$ and $D_i^U(\mathbf{x}_1)\geq0$. Hence $D_i(\mathbf{x}_1;q)\geq0$.

    \textnormal{(ii)} At an interior allocation margin between $S$ and worker $i$, $f_x(a_{S1},x_{S1})-f_x(a_{i1},x_{i1})=\beta D_i(\mathbf{x}_1;q)$. More generally, if a marginal transfer of opportunity from $S$ to $i$ is feasible, the left-hand side is weakly greater than $\beta D_i(\mathbf{x}_1;q)$; if a marginal transfer from $i$ to $S$ is feasible, the inequality is reversed.

\end{proposition}

\begin{proof}
    See Appendix~\ref{app:proof_prop1}.
\end{proof}

Part (i) formalizes the dynamic value of worker opportunity. Additional opportunity can raise future skill and improve the information available for subsequent allocation, and these two channels need not be additively separable. Part (ii) shows how this dynamic value enters the organization's allocation decision: at an interior margin, the marginal current-output cost of shifting opportunity from $S$ to worker $i$ equals the discounted marginal continuation value of doing so. The boundary conditions allow the same framework to accommodate both full reliance on $S$ and more dispersed allocations. Information can be valuable after either favorable or unfavorable performance surprises if it improves subsequent task allocation.

\begin{remark}[Organizational horizon]
    As $\beta$ decreases, the organization places less weight on the dynamic value of worker opportunities; when $\beta=0$, allocation is purely static. A sufficiently short horizon can therefore make heavy reliance on $S$ optimal, although diminishing returns may still favor some dispersion of opportunities.
\end{remark}

\begin{remark}[Future high-performer unavailability]
    Holding $\mathbf{x}_1$ fixed, $\partial D_i(\mathbf{x}_1;q)/\partial q = D_i^U(\mathbf{x}_1)-D_i^A(\mathbf{x}_1)$. Greater future unavailability therefore raises the marginal value of worker opportunity when $D_i^U>D_i^A$, but not in general. This condition is natural when the absence of $S$ makes selecting among substitute workers more consequential.
\end{remark}

\begin{corollary}[Task capacity]
    Because $x_{S1}\leq1$, the share of total related-task capacity allocated to $S$ satisfies $x_{S1}/K\leq1/K$. Thus, the maximum proportional crowd-out generated by a single high-productivity worker decreases with $K$.
\end{corollary}

This result concerns aggregate worker opportunity. An individual-level attenuation result additionally requires that workers compete for a common pool of opportunities within the $K$-slot task family.

The framework motivates three empirical implications. High-performer absence should expand opportunities for other workers, with weaker crowd-out when related-task capacity is larger. Expanded opportunities may affect subsequent productivity through human-capital accumulation and subsequent allocation through information revealed by performance surprises. The counterfactual analysis quantifies the latter channel by comparing allocations with and without the information generated during high-performer absences.

\section{Data} \label{sec:data}

\subsection{Data Sources and Sample Period}
\label{subsec:data-sources}

We combine data from Retrosheet,\footnote{Retrieved between July 31 and August 18, 2026, from \url{https://www.retrosheet.org/}.} Baseball Prospectus,\footnote{Retrieved May 3, 2025, from \url{https://www.baseballprospectus.com/}.} and the Lahman Baseball Database.\footnote{Retrieved February 25, 2025, from \url{http://seanlahman.com/}.} Retrosheet provides detailed game-level records on player participation, starting lineups, defensive assignments, and batting outcomes, which form the basis of our measures of playing opportunities and production, as well as our assignment of exact defensive positions. Baseball Prospectus provides the dates and types of injured-list (IL) placements. The Lahman Baseball Database provides historical All-Star selections and other player- and season-level records used for player identification, completed-season batting measures, career histories, and longer-horizon outcomes.

Our main analysis covers the 1996--2016 MLB seasons and focuses on 15-day and 60-day IL placements. Players placed on these lists were ineligible to return for at least 15 and 60 days, respectively, and were removed from the active roster, allowing teams to use the roster spot for another player.\footnote{During our sample period, MLB officially referred to the injured list as the disabled list (DL); we use the current term IL throughout. MLB introduced a separate 7-day IL for concussions in 2011. We do not use 7-day IL placements to define focal injury events, but account for them when determining whether comparison high performers remain available during an event window.} We end the sample in 2016 because MLB replaced the 15-day list with a 10-day list in 2017, changing the minimum period of short-term unavailability.

The analysis focuses on the eight non-pitcher defensive positions: catcher, first base, second base, third base, shortstop, left field, center field, and right field. We exclude pitchers because starting rotations and relief roles generate a substantially different allocation structure, and designated hitters, who bat without playing a defensive position, because our empirical design defines substitution relationships using defensive positions.

\subsection{Definition of High-Performing Players}
\label{subsec:high-performers}

In the baseline analysis, we use established MLB stars as our empirical measure of high-performing players. Following \cite{call2015stargazing}, star employees are characterized by exceptional performance, high visibility, and relevant social capital. MLB All-Star selection provides a useful empirical analogue to these features: it reflects high performance and confers league-wide visibility and recognition. Although relevant social capital is difficult to observe directly, repeated participation in this prominent league-wide setting captures this dimension more closely than a purely statistical performance threshold. We therefore define a high-performing player in season $t$ as a player who had been selected to the MLB All-Star Game in at least two prior seasons. By construction, this definition uses only information available before the event season. Among unique players who appeared at one of the eight non-pitcher defensive positions during 1996--2016, 9.6 percent satisfy this criterion in at least one season.

Requiring at least two prior All-Star selections helps distinguish established stars from players who experience a single exceptional season. A one-selection threshold risks including temporary standouts, whereas a substantially higher threshold would exclude many players with sustained records of strong performance and substantially reduce the number of usable injury events. Our baseline requirement therefore balances selectivity with empirical support and identifies players whose performance and reputation were already well established before the injury shock.\footnote{As a robustness check, we alternatively define high-performing players using prior-season batting performance. Specifically, we use qualified hitters---those with at least 3.1 plate appearances per scheduled team game (446 plate appearances in 1995 and 502 thereafter)---in the top 10 percent of the prior-season on-base plus slugging (OPS) distribution. OPS is the sum of on-base percentage and slugging percentage and serves as a rate-based measure of batting productivity. The crowd-out estimates remain positive across all four outcomes under this alternative statistical definition.}

\section{Empirical Strategy} \label{sec:empirical}

\subsection{Injury-Based Research Design and Identification}
\label{subsec:injury-design}

Our empirical strategy exploits temporary IL-induced absences of high-performing players in an event-study framework. We treat each eligible IL placement as an event and compare outcomes around the resulting absence with those around the same calendar date for high-performing players in the same season and defensive position on other teams who remain available. Event time is measured in team games relative to the IL start date, with event time 0 defined as the first focal-team game on or after that date.\footnote{In some cases, the focal player appears in a game on the recorded IL start date, as the recorded date may correspond to an injury or administrative event occurring on a game day rather than to the first game of unavailability. Because the high performer remains available in that game, we define event time 0 as the following team game.} For comparison teams, the corresponding pseudo-event begins with the first team game on or after the treated event date.

We use two complementary units of analysis. In the worker-level analyses, we identify potential substitute workers using only information available before the event and hold this worker pool fixed thereafter. In the team-position analysis, we instead allow the set of players supplying the focal position to change, thereby capturing lineup changes, positional switches, and roster adjustments.

The identifying assumption is that, conditional on the comparison design, the timing of high-performer injuries is unrelated to the potential outcomes of the workers who would replace them. A natural threat is that teams with weaker substitutes may rely more heavily on their high performer, increasing his workload and injury risk. Table~\ref{tab:injury-plausibility} examines this possibility by testing whether the quality of the strongest predetermined same-position backup predicts which high performer in a matched risk set experiences the injury. We measure backup quality using both recent pre-event OPS and previous-season OPS. Across these measures, we find little evidence of a systematic relationship between backup quality and injury incidence. This evidence supports treating the injury shocks as plausibly exogenous to the quality of the predetermined substitute pool.

\begin{table}[!htbp]
\centering
\caption{Backup Quality and High-Performer Injury Incidence}
\label{tab:injury-plausibility}
\begin{tabular}{lcc}
\toprule
 & Recent pre-injury & Prior season \\
 & (1) & (2) \\ \midrule
Best-backup performance & 0.002 & 0.008 \\
 & (0.023) & (0.026) \\
R-squared & 0.000 & 0.000 \\
Observations & 2,162 & 1,882 \\
\bottomrule
\end{tabular}

\medskip
\begin{minipage}{0.88\textwidth}
\footnotesize
\textit{Notes:} The dependent variable is an indicator equal to one for the injured side and zero for comparison sides within each matched event risk set. Best-backup performance is the standardized OPS of the strongest predetermined same-position backup, measured using recent pre-event performance in column (1) and prior-season performance in column (2). All specifications include event-risk-set fixed effects. Within each risk set, the injured side receives weight one and comparison sides jointly receive weight one. Standard errors, two-way clustered by focal event and high performer, are in parentheses. $^{*}p<0.10$, $^{**}p<0.05$, $^{***}p<0.01$.
\end{minipage}
\end{table}

\subsection{Crowd-Out of Playing Opportunities}
\label{subsec:crowdout-design}

We first examine whether a high-performing player's absence expands the playing opportunities of incumbent coworkers. Because position serves a different purpose for the high performer and for potential substitutes, we use different information hierarchies to define it. For a high performer, the relevant position is intended to capture the task he was filling for the focal team before the absence. We therefore determine his position from current-season defensive assignments with the focal team prior to the event. If no position can be assigned from these records, we use current-season assignments with other teams and, if still unresolved, information from the previous season. For potential substitute workers, by contrast, we seek to capture their predetermined primary occupation. We therefore use the completed previous season whenever available and use current-season information strictly before the event only when the lagged position cannot be resolved.\footnote{Within each information window, we define a player’s primary defensive position as the position at which he accumulated the most team defensive outs while fielding that position. These are team defensive outs, not individual putouts. Ties are resolved using starts, then appearances, and finally the following positional hierarchy: catcher $>$ shortstop $>$ second base $>$ third base $>$ center field $>$ right field $>$ left field $>$ first base.}

For each treated event, the predetermined worker pool consists of non-star position players affiliated with the focal team who appeared for the team in at least one of the preceding 40 team games and whose predetermined position matches the focal position. We construct comparison workers analogously at each pseudo-event and hold membership in the worker pool fixed thereafter.\footnote{If a comparison high performer subsequently becomes unavailable because of an IL placement, becomes the subject of a focal injury event, or leaves the team, subsequent comparison observations are censored.}

Our primary outcome is plate appearances (PA), a direct measure of batting opportunities. We also examine indicators for appearing in a game in a non-pitcher role, starting a game in any non-pitcher role, and starting at the focal defensive position. We estimate effects over event time $k=-20,\ldots,20$, using $k=-2$ as the reference period. Game $k=-1$ is estimated separately because it may be injury-adjacent. We retain worker-event stacks only when both $k=-2$ and $k=0$ are observed; observations at more distant event times need not form a balanced panel. Stacking each treated event with its comparison units, we estimate
\begin{equation}
Y_{iek}
=
\sum_{\substack{r=-20 \\ r\neq -2}}^{20}
\beta_r
\left[
\mathbbm{1}\{k=r\}\times Treat_{ie}
\right]
+
\alpha_{ie}
+
\lambda_{ek}
+
\varepsilon_{iek},
\label{eq:eventstudy}
\end{equation}
where $Treat_{ie}$ equals one for workers on the treated team, $\alpha_{ie}$ denotes worker-by-event fixed effects, and $\lambda_{ek}$ denotes event-by-relative-game fixed effects. Thus, $\beta_r$ measures the treated--control difference at event time $r$ relative to game $-2$. Within each event, treated workers collectively receive weight one. Comparison units collectively receive weight one, divided equally across comparison units with worker support and then equally among workers within each unit. Standard errors are clustered two ways, by focal event and player.

Motivated by Corollary~1, we also test whether crowd-out is weaker when workers have access to a broader set of closely related task slots. We group left field, center field, and right field into a single outfield task family, while treating catcher and each infield position as separate task families. For this analysis, the task family is defined before assigning focal high performers, comparison high performers, and workers, so that all three are classified symmetrically at the same task-family level. Using the same $-20$ to $+20$ event window as in Figure~\ref{fig:crowdout}, we summarize the dynamic response by comparing a clean pre-event period, covering games $-20$ through $-2$, with vacancy games from 0 through $+20$ while the focal high performer remains unavailable. Injury-adjacent game $-1$ is estimated separately. We then interact the treated-vacancy effect with an outfield indicator, using the same four measures of playing opportunities.

\subsection{Production and Productivity Around High-Performer Absences}
\label{subsec:production-productivity}

We next extend the design beyond the onset of the absence to examine production during the full IL vacancy and after the high performer returns. We define four lifecycle periods: a clean pre-event period covering team games $-5$ through $-2$, the injury-adjacent game $-1$, the full administrative vacancy, and return-relative games 0 through 20. Return game 0 is the first team game on or after the high performer's IL reinstatement date. We stop the treated return window if the original focal high performer becomes unavailable again or is subsequently observed playing for another team. For each comparison unit, we construct a pseudo-return after the same number of team games as the corresponding treated vacancy. We require a complete four-game clean pre-event period and restrict the analysis to treated events with a same-season reinstatement; subsequent return observations need not form a balanced panel.

Let $\tau$ index lifecycle time, where $\tau=P$ denotes the clean pre-event period, $\tau=-1$ the injury-adjacent game, $\tau=V$ the full vacancy, and $\tau=0,\ldots,20$ games relative to reinstatement. For unit $u$ in event $e$, we estimate
\begin{equation}
Y_{eu\tau}
=
\sum_{s\in\{-1,V,0,\ldots,20\}}
\beta_s
\left[
\mathbbm{1}\{\tau=s\}\times Treat_{eu}
\right]
+
\alpha_{eu}
+
\lambda_{e\tau}
+
\varepsilon_{eu\tau},
\label{eq:lifecycle}
\end{equation}
where the clean pre-event period $P$ is the omitted category, $\alpha_{eu}$ denotes unit-by-event fixed effects, and $\lambda_{e\tau}$ denotes event-by-lifecycle-time fixed effects. Count outcomes are expressed per team game within each lifecycle period, while rate outcomes are pooled over the observations in that period. As in the preceding analysis, treated units receive weight one and comparison units jointly receive weight one within each event. Standard errors are clustered two ways, by focal event and team-season.

We apply this specification first at the team-position level to measure the production consequences of the absence while allowing teams to adjust how the focal task is supplied. In each game, we identify all players who field the focal position and attribute their batting outcomes in proportion to their shares of defensive outs recorded at that position. The set of suppliers may therefore change through substitutions, position switches, or roster adjustments. We consider total bases, times on base, runs scored, and OPS. Total bases weight singles, doubles, triples, and home runs by one, two, three, and four bases, respectively; times on base equal hits plus walks and hit-by-pitches. OPS serves as our main rate-based measure of batting productivity.

We then apply the same lifecycle specification to workers defined using the predetermined-worker criteria in Section~\ref{subsec:crowdout-design}, subject to the lifecycle-specific support requirements described above. Worker membership is fixed within each included event, and the focal high performer is excluded. Production is summed across these workers in each game, with nonparticipation contributing zero production. Comparing total production with OPS distinguishes the mechanical output effect of receiving additional opportunities from changes in productivity per unit of opportunity. Under a learning-by-doing mechanism, the additional experience generated during the vacancy may raise worker productivity during the absence and, importantly, after the high performer returns.

\subsection{Information Revelation and Subsequent Allocation} \label{subsec:empirical_information}

A second source of dynamic value from the additional opportunities is information about worker ability. We examine whether performance revealed during the vacancy predicts how teams allocate opportunities afterward. For each predetermined treated worker, we measure prior productivity using OPS over up to his most recent 200 plate appearances observed before the event and compare it with his pooled OPS during the IL vacancy. We define the performance surprise as
\begin{equation}
Surprise_{ie}
=
OPS^{vacancy}_{ie}
-
OPS^{prior}_{ie},
\label{eq:surprise}
\end{equation}
where a positive value indicates that worker $i$ performed better during event $e$ than his recent pre-event record would predict. The signal therefore uses only performance realized during the vacancy relative to information available before the absence.

We relate this signal to subsequent worker outcomes using
\begin{equation}
Y_{ie}
=
\beta Surprise_{ie}
+
\gamma OPS^{prior}_{ie}
+
X_{ie}'\theta
+
\delta_t
+
\phi_p
+
\mu_g
+
\varepsilon_{ie},
\label{eq:signal}
\end{equation}
where $X_{ie}$ includes career plate appearances before the event, the number of prior MLB seasons, and age at the event. The specification includes event-season ($\delta_t$), focal-position ($\phi_p$), and treated-team ($\mu_g$) fixed effects. Workers receive equal weight within each event, and standard errors are clustered two ways, by focal event and player.

We consider four measures of subsequent allocation: plate appearances per game over the same-season post-return period, an indicator for whether the worker's MLB career continues beyond the event season, and plate appearances and starts in the immediately following season. We restrict all four outcomes to temporary-absence episodes in which the original focal high performer returns within the same season, so that subsequent allocation is observed after the original staffing constraint has ended rather than during a persistent shortage. For the latter two outcomes, workers who do not appear in MLB in the following season are assigned zero.\footnote{We additionally consider career-to-date OPS as the prior benchmark and construct analogous performance surprises using total bases and times on base per plate appearance.}

\section{Results} \label{sec:result}

This section presents the empirical results corresponding to the three implications of the conceptual framework: crowd-out and team production, human-capital accumulation, and information discovery.

\subsection{Crowd-Out and Team Production} \label{subsec:crowd-out}

\begin{figure}[!htbp]
\centering
\includegraphics[width=0.96\textwidth]{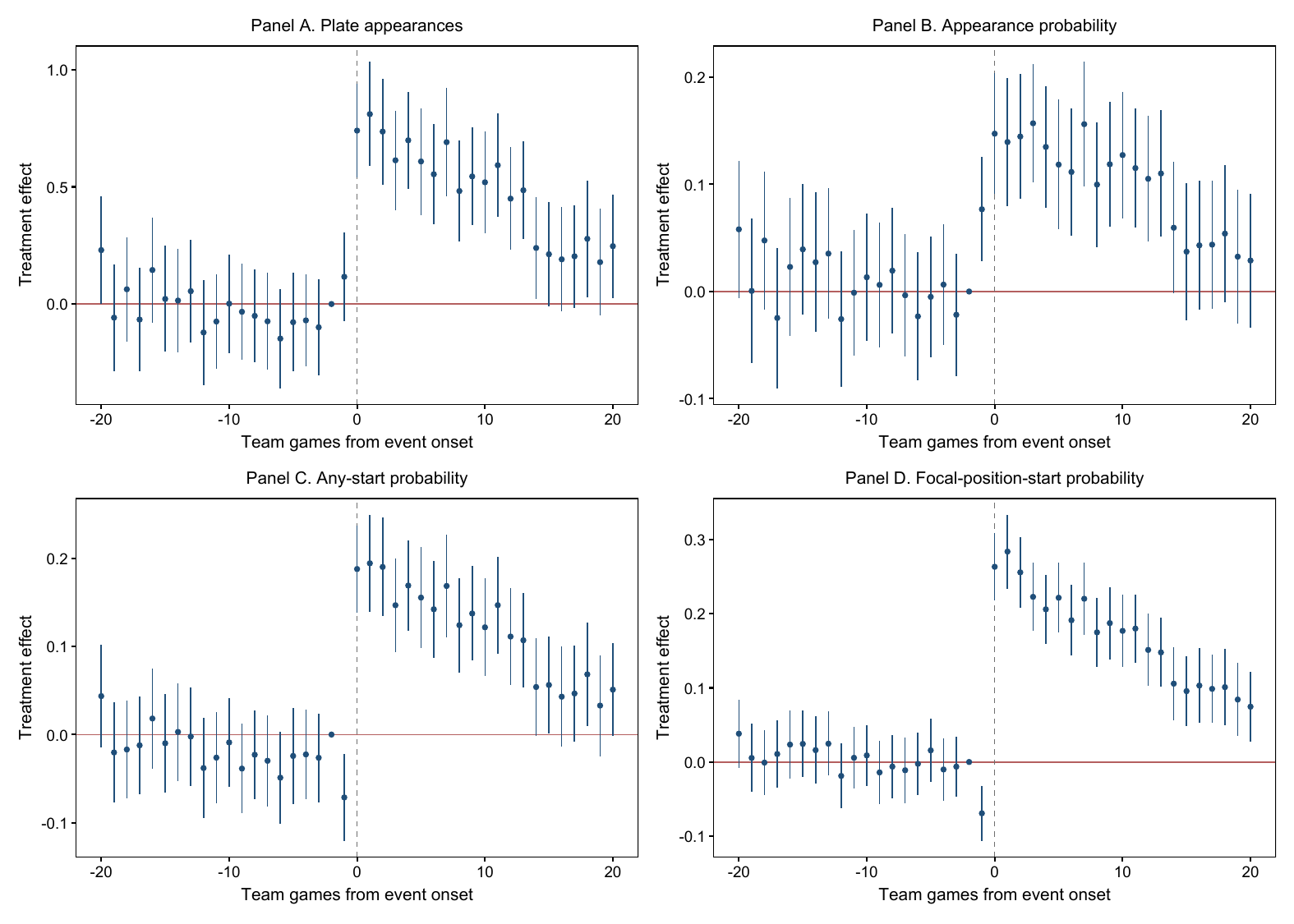}
\caption{Crowd-Out of Playing Opportunities}
\label{fig:crowdout}

\medskip
\begin{minipage}{0.94\textwidth}
\footnotesize
\textit{Notes:} Each panel reports event-study estimates for the fixed pool of pre-existing same-team, same-position workers. Event time is measured in focal-team games from event onset; game $-2$ is the omitted category, and injury-adjacent game $-1$ is estimated separately. Models include worker-by-event and event-by-relative-game fixed effects and use event-balanced worker weights. The sample contains 392 focal events and 2,943 worker-event units. Joint Wald tests of $H_0:\beta_{-20}=\cdots=\beta_{-3}=0$ yield $p$-values of 0.108, 0.110, 0.173, and 0.521 for Panels A--D, respectively. Pointwise 95 percent confidence intervals use standard errors two-way clustered by focal event and player.
\end{minipage}
\end{figure}

Figure~\ref{fig:crowdout} shows how playing opportunities for pre-existing same-team, same-position coworkers change around the onset of a high-performing player's absence. These workers are defined using information available before the absence, so the composition of the worker group does not respond to subsequent playing time. We measure playing opportunities using plate appearances and indicators for appearing in a game, making any start, and starting at the focal position. Across all four measures of playing opportunities, we find a similar pattern. Opportunities increase significantly when the high performer becomes unavailable, and the effects persist for roughly the first 12 team games following onset. This duration is broadly consistent with the number of games typically played during the 15-day minimum IL period.\footnote{As a robustness check, we alternatively define high-performing players as qualified hitters in the top 10 percent of the prior-season OPS distribution. The crowd-out estimates remain positive across all four outcomes under this alternative statistical definition.} By contrast, the pre-onset estimates are close to zero and statistically insignificant. The event-time patterns therefore provide clear evidence that high-performing workers crowd out the playing opportunities of their same-position coworkers.

The conceptual framework further predicts that crowd-out should be weaker when workers can compete for a broader set of related task opportunities. We examine this implication by exploiting differences across positions. Because positional switching is relatively common among left field, center field, and right field compared with other defensive positions, we group the three outfield positions into a single task family. We then compare outfielders with catcher and the four infield positions, which are treated as separate positions. Table~\ref{tab:task-capacity} reports these heterogeneous crowd-out effects by position group. Across all four measures of playing opportunities, the increase following a focal absence is approximately 57--61 percent smaller for outfielders, and the interaction terms are statistically significant across outcomes. Thus, crowd-out is systematically weaker when workers have access to a broader set of closely related task opportunities.

\begin{table}[!htbp]
\centering
\caption{Task Capacity and Crowd-Out}
\label{tab:task-capacity}
\begin{adjustbox}{max width=\textwidth}
\begin{tabular}{lcccc}
\toprule
 & PA & Appearance & Any start & \shortstack{Focal-position\\start} \\
 & (1) & (2) & (3) & (4) \\ \midrule
Focal absence & 0.768*** & 0.152*** & 0.201*** & 0.236*** \\
 & (0.081) & (0.021) & (0.020) & (0.019) \\
Focal absence $\times$ Outfield & -0.440*** & -0.092*** & -0.121*** & -0.143*** \\
 & (0.088) & (0.023) & (0.021) & (0.020) \\
R-squared & 0.462 & 0.405 & 0.431 & 0.418 \\
Observations & 730,809 & 730,809 & 730,809 & 730,809 \\
\bottomrule
\end{tabular}

\end{adjustbox}

\medskip
\begin{minipage}{0.96\textwidth}
\footnotesize
\textit{Notes:} The table compares treated--control vacancy effects on worker playing opportunities between single-seat task groups and the outfield family, which combines left field, center field, and right field before assigning focal high performers, comparison high performers, and workers to task families. The clean pre-event period covers games $-20$ through $-2$; injury-adjacent game $-1$ and its interaction with the outfield indicator are estimated separately but not reported. The vacancy period covers games 0 through $+20$ while the focal high performer remains unavailable. ``Focal absence'' is the vacancy effect for non-outfield task groups, and the interaction gives the differential effect for the outfield family. Models use the fixed predetermined-worker pool, event-balanced worker weights, worker-by-stack fixed effects, and event-by-phase fixed effects. Standard errors, two-way clustered by focal event and player, are in parentheses. $^{*}p<0.10$, $^{**}p<0.05$, $^{***}p<0.01$.
\end{minipage}
\end{table}

\begin{figure}[!htbp]
\centering
\includegraphics[width=0.96\textwidth]{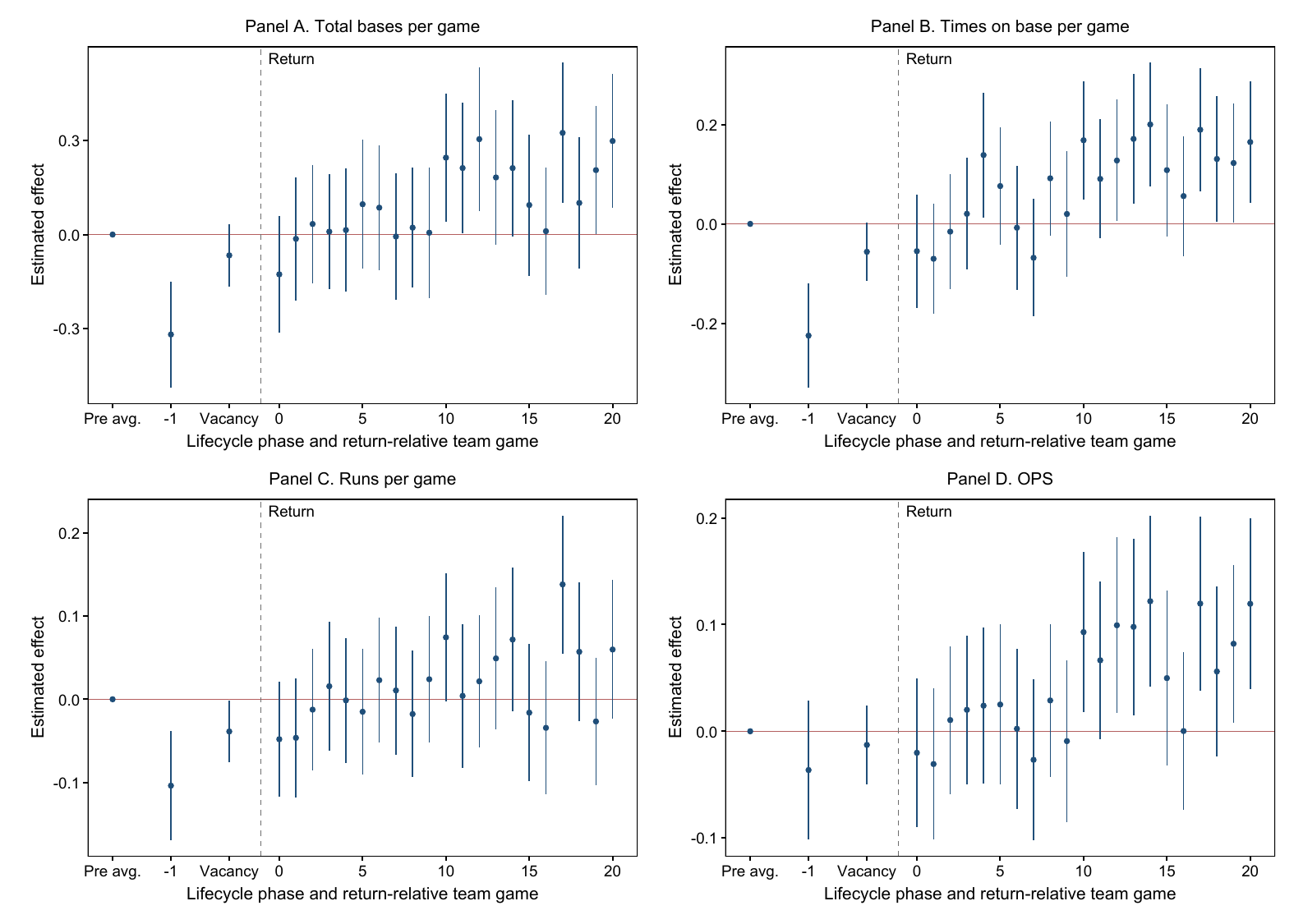}
\caption{Focal-Position Production Around High-Performer Absences}
\label{fig:position-lifecycle}

\medskip
\begin{minipage}{0.94\textwidth}
\footnotesize
\textit{Notes:} The unit is the focal team-position. Batting outcomes are attributed to players who field the focal position in proportion to their shares of defensive outs at that position. The omitted category is the average over event-time games $-5$ through $-2$, and injury-adjacent game $-1$ is reported separately. The vacancy runs from event onset through the game immediately preceding return; return game 0 is the first focal-team game on or after the high performer's IL reinstatement date. The sample contains 485 focal events and requires complete clean-pre and vacancy support and a same-season return. Models include event-unit and event-by-lifecycle-time fixed effects and use event-balanced unit weights. Pointwise 95 percent confidence intervals use standard errors two-way clustered by focal event and team-season. Count outcomes are measured per team game; OPS is pooled from the underlying batting components.
\end{minipage}
\end{figure}

We next examine whether this crowd-out of individual opportunities has consequences for team production. We treat the team-position as the smallest relevant unit of team production and trace production at the position affected by the high performer's absence. Unlike the worker-level analysis above, we do not fix the set of players in advance. Teams can respond to an absence by changing lineups, shifting players across positions, or altering the roster, and allowing the set of focal-position suppliers to change captures these dynamic replacement responses. Figure~\ref{fig:position-lifecycle} reports total bases, times on base, runs, and OPS. During the vacancy, point estimates are modestly negative across all four outcomes, with some evidence of declines in runs and times on base but little evidence of a broad decline in team-position production. After the focal player returns, the estimates generally turn positive, with some becoming statistically significant at later horizons. Overall, the absence is associated with only a modest negative short-run pattern in team-position production, followed by a slight improvement after the focal player returns. This pattern does not appear to be driven primarily by teams bringing in established MLB players from outside the focal team.\footnote{Table~\ref{tab:appendix-external-replacement} shows that external MLB movers appear in 85 of 485 treated events (17.53\%) and account for 4.56--4.79\% of focal-position opportunities and production during the vacancy and 3.06--3.34\% over the first 20 games after return. These movers are typically experienced MLB players rather than previously untested entrants.}

\subsection{Human-Capital Accumulation} \label{subsec:human-capital}

One potential source of dynamic value from the additional opportunities documented above is learning by doing: greater task exposure may raise workers' subsequent productivity. We therefore examine whether workers who receive additional opportunities during a high performer's absence become more productive per unit of opportunity.

\begin{figure}[!htbp]
\centering
\includegraphics[width=0.96\textwidth]{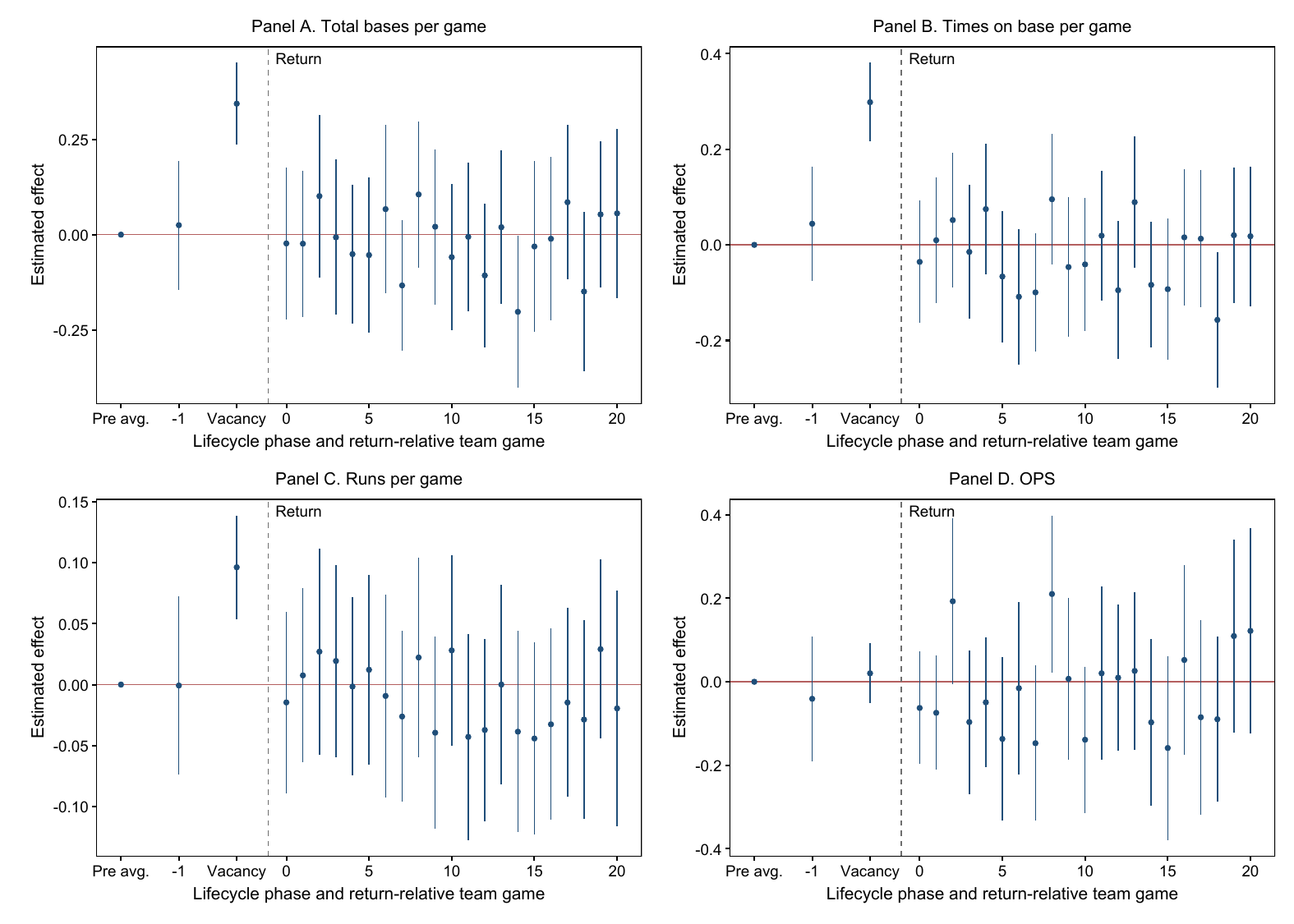}
\caption{Predetermined-Worker Production and Productivity}
\label{fig:worker-lifecycle}

\medskip
\begin{minipage}{0.94\textwidth}
\footnotesize
\textit{Notes:} Outcomes aggregate production by the fixed pool of pre-event same-team, same-position workers and exclude the focal high performer. The omitted category is the average over event-time games $-5$ through $-2$, and injury-adjacent game $-1$ is reported separately. The vacancy runs from event onset through the game immediately preceding return; return game 0 is the first focal-team game on or after the high performer's IL reinstatement date. The sample contains 321 focal events and requires complete clean-pre and vacancy support and a same-season return. Models include event-unit and event-by-lifecycle-time fixed effects and use event-balanced unit weights. Pointwise 95 percent confidence intervals use standard errors two-way clustered by focal event and team-season. Count outcomes are measured per team game; OPS is pooled from the worker-pool batting components.
\end{minipage}
\end{figure}

Figure~\ref{fig:worker-lifecycle} examines this possibility using a predetermined worker pool constructed from pre-event information, as in the crowd-out analysis. The figure reports both total production and rate-based productivity before, during, and after the vacancy. During the vacancy, total production increases significantly, consistent with workers receiving more playing opportunities. By contrast, OPS, a rate-based measure of batting productivity, shows no significant increase. After the focal player returns, total production moves back toward its pre-absence level, and OPS shows no systematic or sustained improvement. Thus, although the additional opportunities generate more output while workers receive them, we find little evidence that these additional opportunities are accompanied by a sustained increase in rate-based productivity.\footnote{As an additional test, Table~\ref{tab:appendix-duration} examines whether longer absences, which provide greater scope for additional task exposure, are associated with larger subsequent improvements in rate productivity. We find little evidence of such a pattern for OPS, total bases per plate appearance, or times on base per plate appearance.}

Taken together, these results provide little evidence that the additional task experience generated by high-performer absences raises subsequent batting productivity. The human-capital channel therefore receives limited empirical support in our setting.

\subsection{Information Discovery and Subsequent Allocation} \label{subsec:information}

A second potential source of dynamic value from productive opportunities is information discovery: observing workers in meaningful tasks may reveal information about their ability that can improve subsequent allocation decisions. We test this implication by examining whether performance surprises during the vacancy predict how workers are allocated opportunities afterward.

Table~\ref{tab:performance-signals} shows a strong relationship between better-than-expected vacancy performance and subsequent allocation. A (+0.100) OPS surprise is associated with 0.160 additional plate appearances per game after the focal player returns and a 3.3 percentage point increase in the probability of continuing an MLB career beyond the event season. The same positive pattern appears in next-season plate appearances and starts. Thus, workers who perform better than their prior records would predict subsequently receive more playing opportunities and are more likely to remain in MLB.

\begin{table}[!htbp]
\centering
\caption{Performance Surprises and Subsequent Worker Allocation}
\label{tab:performance-signals}
\begin{adjustbox}{max width=\textwidth}
\begin{tabular}{lcccc}
\toprule
 & \shortstack{Post-return\\PA/game} & \shortstack{Career\\continuation} & \shortstack{Next-season\\PA} & \shortstack{Next-season\\starts} \\
 & (1) & (2) & (3) & (4) \\ \midrule
Performance surprise (+0.100 OPS) & 0.160*** & 0.033*** & 14.746*** & 3.158*** \\
 & (0.029) & (0.008) & (3.788) & (0.868) \\
R-squared & 0.331 & 0.300 & 0.314 & 0.305 \\
Observations & 430 & 430 & 430 & 430 \\
\bottomrule
\end{tabular}

\end{adjustbox}

\medskip
\begin{minipage}{0.96\textwidth}
\footnotesize
\textit{Notes:} The worker-level performance surprise is vacancy OPS minus OPS over up to the worker's most recent 200 pre-event plate appearances; coefficients are scaled to a +0.100 OPS surprise. Outcomes are same-season post-return PA per game, an indicator for MLB career continuation beyond the event season, and PA and starts in the immediately following MLB season. The sample is restricted to events in which the focal high performer returns within the same season. The same-season post-return window ends if the original focal high performer becomes unavailable again because of a subsequent injury. This censoring applies only to post-return PA per game; career continuation and next-season outcomes are measured over their respective later horizons. Models control for pre-event OPS, pre-event career PA, prior MLB seasons, and age and include event-season, focal-position, and treated-team fixed effects. Workers receive equal weight within each event. Standard errors, two-way clustered by focal event and player, are in parentheses. $^{*}p<0.10$, $^{**}p<0.05$, $^{***}p<0.01$.
\end{minipage}
\end{table}

Because OPS is a rate-based measure, performance surprises based on only a small number of plate appearances may contain substantial noise. Table~\ref{tab:signal-exposure} therefore progressively restricts the sample to workers with at least 5, 10, or 20 plate appearances during the vacancy. The estimated relationships remain broadly similar across these restrictions, indicating that the main results are not driven by performance measures based on very limited exposure.\footnote{The results are also robust to alternative constructions of the performance signal. Table~\ref{tab:appendix-signals} uses career-to-date performance rather than recent performance as the prior benchmark and constructs analogous surprises using total bases and times on base per plate appearance, with qualitatively similar results.}

\begin{table}[!htbp]
\centering
\caption{Robustness to Vacancy-Period Exposure Thresholds}
\label{tab:signal-exposure}
\begin{adjustbox}{max width=\textwidth}
\begin{tabular}{lcccc}
\toprule
 & \shortstack{Post-return\\PA/game} & \shortstack{Career\\continuation} & \shortstack{Next-season\\PA} & \shortstack{Next-season\\starts} \\
 & (1) & (2) & (3) & (4) \\ \midrule
Unrestricted (N = 430) & 0.160*** & 0.033*** & 14.746*** & 3.158*** \\
 & (0.029) & (0.008) & (3.788) & (0.868) \\
Vacancy PA \ensuremath{\geq} 5 (N = 412) & 0.163*** & 0.038*** & 15.200*** & 3.199*** \\
 & (0.034) & (0.008) & (4.659) & (1.078) \\
Vacancy PA \ensuremath{\geq} 10 (N = 395) & 0.202*** & 0.044*** & 18.903*** & 3.917*** \\
 & (0.029) & (0.009) & (5.028) & (1.196) \\
Vacancy PA \ensuremath{\geq} 20 (N = 352) & 0.247*** & 0.027*** & 14.230** & 2.852** \\
 & (0.038) & (0.010) & (5.999) & (1.411) \\
\bottomrule
\end{tabular}

\end{adjustbox}

\medskip
\begin{minipage}{0.96\textwidth}
\footnotesize
\textit{Notes:} Each row re-estimates the Table~\ref{tab:performance-signals} relationship after imposing the indicated minimum number of vacancy-period PA. The performance signal, outcomes, controls, fixed effects, and equal worker weighting within event are otherwise unchanged. Coefficients are scaled to a +0.100 OPS surprise. Standard errors, two-way clustered by focal event and player, are in parentheses. $^{*}p<0.10$, $^{**}p<0.05$, $^{***}p<0.01$.
\end{minipage}
\end{table}

Taken together, these results indicate that information revealed during temporary reallocations is reflected in subsequent personnel and playing-time decisions. Combined with the limited evidence of productivity gains in Section~\ref{subsec:human-capital}, the results provide stronger support for information discovery than for human-capital accumulation as the main source of dynamic value from the additional opportunities generated by high-performer absences.

\section{Counterfactual} \label{sec:counterfactual}

The results in Section~\ref{subsec:information} suggest that temporary high-performer absences generate information that is reflected in subsequent allocation decisions. We now quantify the potential production value of this information. Specifically, we ask how much future production could improve if teams used the information revealed during the vacancy when allocating a fixed set of playing opportunities, and compare this benefit with the production consequences of the absence that generated the information.

For the counterfactual, we use the performance-surprise construction described in Section~\ref{subsec:empirical_information}, expressed in total bases per plate appearance (TB/PA) so that the signal maps directly into our production measure. Because vacancy-period performance is noisy, we allow only a fraction of the observed surprise to update the worker's prior assessment. Let $p_{ie}^{prior}$ and $p_{ie}^{vacancy}$ denote worker $i$'s pre-event and vacancy-period TB/PA in event $e$. The updated assessment is
\begin{equation}
\widetilde{p}_{ie}(\rho)
=
p_{ie}^{prior}
+
\rho
\left(
p_{ie}^{vacancy}
-
p_{ie}^{prior}
\right),
\end{equation}
where $\rho$ is a common, non-estimated sensitivity parameter governing the weight placed on the vacancy-period signal. We report results for $\rho=0.25$, $0.50$, and $0.75$.\footnote{Workers with no plate appearances during the vacancy generate no new performance signal and therefore retain their prior assessment.}

We then compare two assignments of the same fixed amount of additional playing opportunity to the same predetermined worker pool. The prior-information assignment, $P$, selects the worker with the highest pre-event assessment, whereas the informed assignment, $I$, selects the worker with the highest updated assessment. Both assignments are evaluated using the updated productivity measure, so their difference isolates the value of using the newly revealed information when allocating opportunities. The information benefit for event $e$ is the same fixed amount of additional playing opportunity, $M$, multiplied by the difference in updated productivity between the workers selected under $I$ and $P$.

To compare this information benefit with the production consequence of generating it, we first construct an event-level realized production effect as the treated pre-to-vacancy change in focal-position total bases per game minus the corresponding weighted-average change among matched controls, using the same comparison weights as in the main design. Rather than using this noisy realized effect directly, we predict it from pre-onset characteristics. The predictors include pre-event production by the focal high performer, predetermined workers, and other pre-existing focal-position suppliers, together with the number of suppliers, season, and focal position. Valid Figure~\ref{fig:position-lifecycle} training events receive leave-one-event-out predictions, whereas external counterfactual target events receive predictions from the model estimated on the full valid Figure~\ref{fig:position-lifecycle} training population. We estimate one additive alignment constant from that Figure~\ref{fig:position-lifecycle} calibration population and apply the same constant to both training-target and external-target predictions. The final target population is therefore broader than the training population. Let $\widehat{C}_e$ denote the resulting predicted production effect. Our net production value is therefore
\begin{equation}
N_e(\rho)
=
\widehat{C}_e
+
M
\left[
\widetilde{p}_{I,e}(\rho)
-
\widetilde{p}_{P,e}(\rho)
\right].
\end{equation}
Both components are expressed in total bases per focal-star-loss team game. The first captures the predicted production effect of one current vacancy-period game, while the second captures the potential information benefit in one future game in which the relevant opportunities must again be allocated without the key worker. The measure is therefore a per-game production value rather than a spell-level or lifetime welfare calculation.

Table~\ref{tab:counterfactual} summarizes the counterfactual results. The mean predicted realized production effect is approximately $-0.034$ total bases per focal-star-loss team game. The average information benefit is positive and increases with $\rho$, from approximately $0.001$ at $\rho=0.25$ to $0.012$ at $\rho=0.75$, while the mean net value remains modestly negative under each calibration.

\begin{table}[!htbp]
\centering
\caption{Counterfactual Production Costs and Information Benefits}
\label{tab:counterfactual}
\begin{tabular}{lccc}
\toprule
 & $\rho=0.25$ & $\rho=0.50$ & $\rho=0.75$ \\
 & (1) & (2) & (3) \\ \midrule
\multicolumn{4}{l}{\textit{Event-level production components}} \\
Mean predicted realized effect & -0.034 & -0.034 & -0.034 \\
Mean information benefit & 0.001 & 0.006 & 0.012 \\
Mean net production value & -0.033 & -0.028 & -0.021 \\
Median net production value & 0.112 & 0.116 & 0.117 \\
10th percentile of net production value & -1.151 & -1.149 & -1.149 \\
90th percentile of net production value & 0.954 & 0.958 & 0.961 \\
Share with net production value above zero & 0.551 & 0.554 & 0.556 \\
Share with zero information benefit & 0.944 & 0.872 & 0.842 \\
\bottomrule
\end{tabular}

\medskip
\begin{minipage}{0.94\textwidth}
\footnotesize
\textit{Notes:} Entries summarize event-level production effects and information benefits for the final counterfactual target population. All components are measured in total bases per focal-star-loss team game. Valid Figure~\ref{fig:position-lifecycle} training events receive leave-one-event-out predictions of matched event-level focal-position production changes; external target events receive predictions from the model estimated on the full valid Figure~\ref{fig:position-lifecycle} training population. One additive alignment constant estimated from the Figure~\ref{fig:position-lifecycle} calibration population is applied to both sets of predictions. The information benefit applies the stated value of $\rho$ to the event-specific future assignment comparison. Net production value is the sum of the predicted realized effect and information benefit.
\end{minipage}
\end{table}

\begin{figure}[!htbp]
\centering
\includegraphics[width=0.82\textwidth]{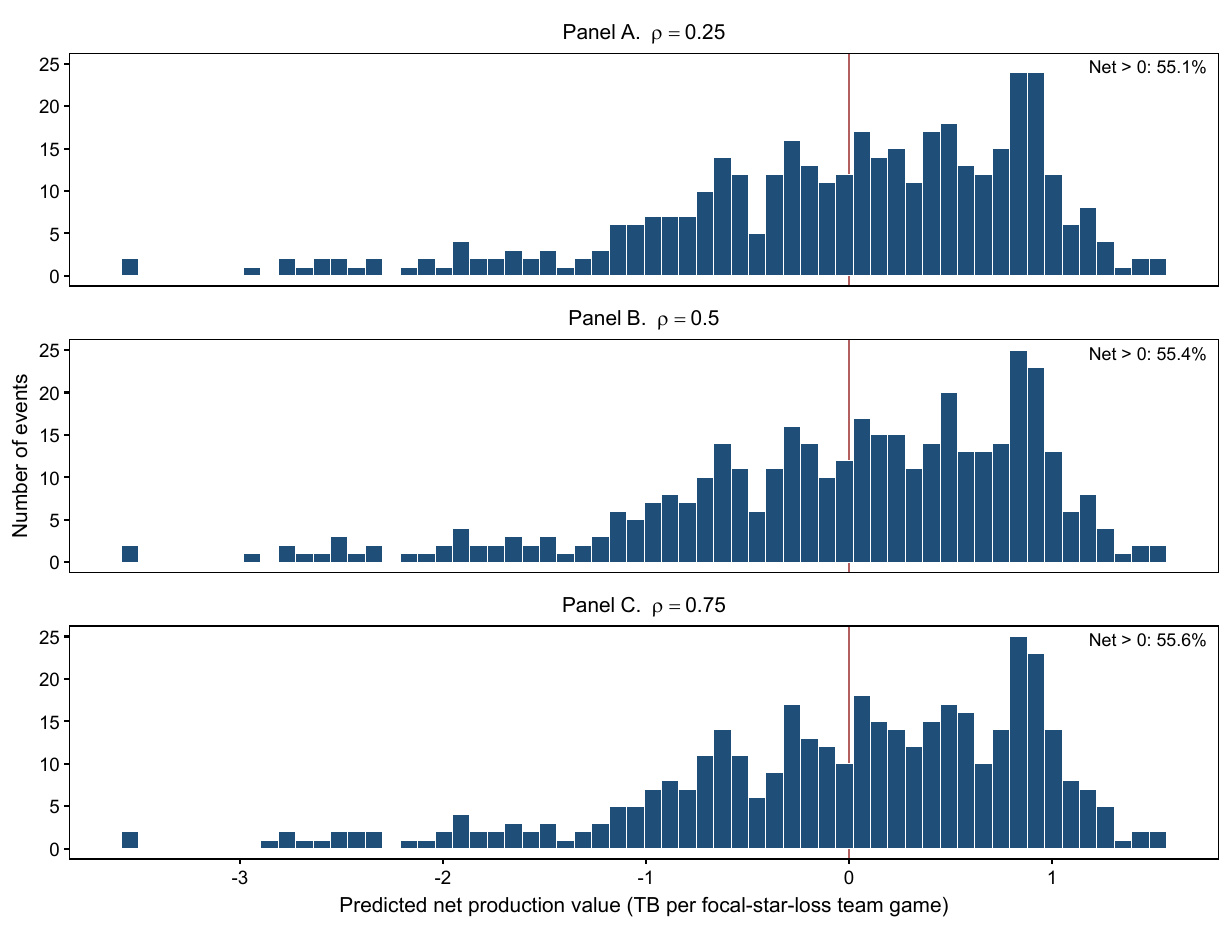}
\caption{Cross-Event Distribution of Net Production Value}
\label{fig:counterfactual-distribution}

\medskip
\begin{minipage}{0.84\textwidth}
\footnotesize
\textit{Notes:} Each panel shows the cross-event distribution of net production value, defined as the predicted realized focal-position production effect plus the event-specific information benefit under the indicated value of $\rho$. The unit is total bases per focal-star-loss team game. Panels use the same event sample, horizontal scale, and 60 equal-width histogram bins. The vertical line marks zero, and each panel reports the share of events with positive net production value. The sample contains 392 events.
\end{minipage}
\end{figure}

\clearpage
Figure~\ref{fig:counterfactual-distribution} shows substantial heterogeneity across events. The median net value is positive under all three values of $\rho$, whereas a relatively small number of events with large predicted production losses generate a pronounced negative tail. The net value is positive in slightly more than half of the events. New information changes the preferred worker only in a minority of cases, but when it does, it creates additional future production value. Thus, the production cost of temporary high-performer absence can be substantial in some cases, while the information generated by the resulting reallocation provides an offsetting source of value in others.

\section{Conclusion} \label{sec:conclusion}

It is often natural for organizations to concentrate important tasks on workers who are already known to perform them well. Yet doing so may carry a less visible cost. Productive opportunities generate not only current output but also human capital and information about the workers who receive them. This paper studies this trade-off using temporary injuries to high-performing Major League Baseball players, which disrupt teams' usual allocation of playing opportunities and force them to rely more heavily on other workers.

We find clear evidence of crowd-out: when a high-performing player becomes unavailable, same-team, same-position coworkers receive more playing opportunities. The additional experience created by these absences provides little evidence of subsequent productivity gains, offering little support for human-capital accumulation as the main source of dynamic value. Instead, the results point to an informational channel. Performance relative to what could have been expected from workers' prior records predicts subsequent employment and playing opportunities, suggesting that teams use the information revealed during temporary reallocations when making later assignment decisions. At the team-position level, production does not fall sharply during the absence and is somewhat higher after the high performer returns. Our counterfactual analysis further shows that information revealed during temporary reallocations can improve future production through better assignment, providing an offsetting source of value when the disruption is costly.

The broader implication is that task allocation shapes not only current production but also what organizations learn about their workers. An organization may allocate tasks efficiently given what it currently knows while learning too little about workers who have not yet been tested in consequential roles. This tension is especially relevant when important tasks are scarce, worker ability is imperfectly known, and performance on the task provides information that is difficult to obtain otherwise. In such environments, repeatedly assigning important tasks to proven performers can protect current production while slowing the discovery of capable alternatives. The same logic may apply when only a small number of physicians perform complex procedures, lawyers or consultants lead important engagements, or employees take responsibility for critical projects.

Our findings do not imply that organizations should routinely take tasks away from their best performers. The disruptions we study create a window in which alternative workers may receive repeated opportunities, so our results do not imply that a few isolated assignments would generate comparable information. Moreover, observed production losses may understate the full cost of experimentation when organizations use costly hiring, transfers, or other adjustments to replace a high performer. Our counterfactual should therefore be interpreted as a comparison of predicted production consequences and informational gains, rather than a complete welfare calculation. Even with these qualifications, evaluating task allocation solely by immediate output may overlook the value of giving less-proven workers meaningful opportunities to reveal their ability. More generally, organizations choosing how heavily to rely on proven talent face a dynamic trade-off between exploiting what they already know and learning who else can perform important tasks well.

\bibliographystyle{ecca}
\bibliography{superstar}

\clearpage
\appendix
\setcounter{section}{0}
\renewcommand{\thesection}{\Alph{section}}
\renewcommand{\thesubsection}{\Alph{section}.\arabic{subsection}}
\setcounter{table}{0}
\renewcommand{\thetable}{\thesection\arabic{table}}

\section*{Appendix}

\section{Proof of Proposition 1}
\label{app:proof_prop1}

\begin{proof}
\textit{Part (i).}
Fix worker $i$ and $\mathbf{x}_{-i,1}$, where $\mathbf{x}_{-i,1}$ denotes the period-1 opportunity allocations of all non-$S$ workers other than $i$, and consider $x'_{i1}>x_{i1}$. The argument applies separately to each availability state $r\in\{A,U\}$.
First, hold the information structure fixed. The posterior mean of worker $i$'s period-2 skill after any signal realization $z$ is $(1-\delta)E[s_{i1}\mid z]+g_i(x_{i1})$. Since $g_i'(x)\geq0$, increasing opportunity from $x_{i1}$ to $x'_{i1}$ raises this posterior mean by $g_i(x'_{i1})-g_i(x_{i1})\geq0$. Since $f_a>0$, expected period-2 output under any given feasible allocation is therefore weakly higher.

Second, hold the human-capital term fixed and consider only the change in information. By the Blackwell monotonicity condition, the information structure under $x'_{i1}$ is weakly more informative than that under $x_{i1}$. Any decision rule under the less informative structure can be reproduced by garbling the more informative signal. Hence additional information cannot reduce optimal expected period-2 output.

Therefore, $V^r(x'_{i1},\mathbf{x}_{-i,1}) \geq V^r(x_{i1},\mathbf{x}_{-i,1})$ for $r\in\{A,U\}$. Wherever the continuation values are differentiable, it follows that 
\[
    D_i^A(\mathbf{x}_1)\geq0\quad \text{and}\quad D_i^U(\mathbf{x}_1)\geq0,
\]
and hence
\[
    D_i(\mathbf{x}_1;q)=(1-q)D_i^A(\mathbf{x}_1)+qD_i^U(\mathbf{x}_1)\geq0.
\]

\medskip
\noindent
\textit{Part (ii).}
Fix worker $i$ and hold $\mathbf{x}_{-i,1}$ fixed. The capacity
constraint then determines $x_{S1}$ residually as $x_{i1}$ varies.
At an interior allocation between $S$ and worker $i$, the first-order
condition is
\[
    f_x(a_{S1},x_{S1})
    -
    f_x(a_{i1},x_{i1})
    =
    \beta D_i(\mathbf{x}_1;q).    
\]
At a boundary, the corresponding one-sided first-order condition applies.
If a marginal transfer from $S$ to $i$ is feasible, then
$f_x(a_{S1},x_{S1})-f_x(a_{i1},x_{i1})
\geq \beta D_i(\mathbf{x}_1;q)$.
If a marginal transfer from $i$ to $S$ is feasible, the inequality is
reversed.
\end{proof}

\section{Additional Results and Robustness Checks}
\label{app:additional-results}

This appendix presents additional analyses and robustness checks referenced in the main text.

\subsection{Alternative Definition: Top 10\% of Prior-Season OPS}
\label{app:alternative-high-performers}

\begin{table}[!htbp]
\centering
\caption{High Performers Defined by the Top 10\% of Prior-Season OPS}
\label{tab:appendix-high-achiever}
\begin{adjustbox}{max width=\textwidth}
\begin{tabular}{lcccc}
\toprule
 & PA & Appearance & Any start & \shortstack{Focal-position\\start} \\
 & (1) & (2) & (3) & (4) \\ \midrule
Focal absence & 0.290* & 0.026 & 0.077* & 0.185*** \\
 & (0.174) & (0.044) & (0.043) & (0.042) \\
R-squared & 0.506 & 0.466 & 0.476 & 0.385 \\
Observations & 7,088 & 7,088 & 7,088 & 7,088 \\
\bottomrule
\end{tabular}

\end{adjustbox}

\medskip
\begin{minipage}{0.96\textwidth}
\footnotesize
\textit{Notes:} Focal and comparison high performers are qualified hitters in the top 10 percent of the completed prior-season OPS distribution. Workers are not themselves in this alternative high-performer category and otherwise follow the predetermined same-team, same-position eligibility rules. The estimation sample contains 48 events, 80 natural controls, and 187 worker stacks. The clean pre-event period covers games $-20$ through $-2$; injury-adjacent game $-1$ is estimated separately but not reported. The vacancy period covers games 0 through $+20$ while the focal high performer remains unavailable. Entries report treated--control vacancy effects on worker playing opportunities using the fixed predetermined-worker pool, event-balanced worker weights, worker-by-stack fixed effects, and event-by-phase fixed effects. Standard errors, two-way clustered by focal event and player, are in parentheses. $^{*}p<0.10$, $^{**}p<0.05$, $^{***}p<0.01$.
\end{minipage}
\end{table}

\clearpage
\subsection{External Replacement During High-Performer Absences}
\label{app:external-replacement}

\begin{table}[!htbp]
\centering
\caption{External MLB Movers During High-Performer Absences}
\label{tab:appendix-external-replacement}
\begingroup
\let\originalvspace\vspace
\renewcommand{\vspace}[1]{\par\originalvspace{#1}}
\textit{Panel A. Prevalence of External MLB Movers}\par\smallskip
\begin{tabular}{lr}
\toprule
Measure & Value \\
\midrule
Treated events & 485 \\
Events with any post-onset entrant & 288 \\
Events with \ensuremath{\geq}1 external mover & 85 \\
Share of treated events with \ensuremath{\geq}1 external mover & 17.53\% \\
External-mover player-events & 100 \\
Unique external-mover players & 93 \\
\bottomrule
\end{tabular}
\vspace{0.8em}
\textit{Panel B. Share of Focal-Position Opportunity and Production}\par\smallskip
\begin{tabular}{lcc}
\toprule
 & \multicolumn{2}{c}{Share of focal-position total} \\
Measure & Vacancy & Return 0:+20 \\
\midrule
Defensive outs & 4.70\% & 3.28\% \\
Total bases & 4.63\% & 3.14\% \\
Times on base & 4.79\% & 3.06\% \\
Runs & 4.56\% & 3.34\% \\
\bottomrule
\end{tabular}
\vspace{0.8em}
\textit{Panel C. Pre-Focal-Team MLB Experience}\par\smallskip
\begin{tabular}{lrr}
\toprule
Measure & Value & N \\
\midrule
Mean prior MLB seasons & 7.2 & 89 \\
Mean prior career PA & 2,667.4 & 89 \\
Mean prior-season PA & 368.6 & 89 \\
Mean prior-season OPS & 0.724 & 89 \\
Share with prior All-Star experience & 24.7\% & 93 \\
Share established regular & 32.6\% & 89 \\
\bottomrule
\end{tabular}

\endgroup

\medskip
\begin{minipage}{0.96\textwidth}
\footnotesize
\textit{Notes:} A post-onset entrant is a player who supplies the focal position after event onset, is neither the focal high performer nor a predetermined worker, and whose first game for the focal team occurs at or after event onset. An external mover is a post-onset entrant who had appeared for another MLB team earlier in the same season before first appearing for the focal team. Panel A reports event prevalence; Panel B reports external movers' shares of defensive outs and batting production among all realized focal-position suppliers during the vacancy and return games 0 through 20. The return window ends if the original focal high performer becomes unavailable again or is subsequently observed playing for another team. Panel C describes movers' MLB experience before joining the focal team. We classify a player as an established regular if he recorded at least 446 plate appearances in 1995 or 502 in later seasons.
\end{minipage}
\end{table}

\clearpage
\subsection{Absence Duration and Subsequent Productivity}
\label{app:absence-duration}

\begin{table}[!htbp]
\centering
\caption{Absence Duration and Subsequent Worker-Pool Productivity}
\label{tab:appendix-duration}
\begin{tabular}{lccc}
\toprule
 & OPS & TB/PA & TOB/PA \\
 & (1) & (2) & (3) \\ \midrule
Vacancy duration (1 SD) & -0.041 & -0.033 & -0.009 \\
 & (0.045) & (0.031) & (0.015) \\
R-squared & 0.596 & 0.588 & 0.601 \\
Observations & 5,728 & 5,743 & 5,743 \\
\bottomrule
\end{tabular}

\medskip
\begin{minipage}{0.90\textwidth}
\footnotesize
\textit{Notes:} Entries report the treated post-return productivity differential associated with a one-standard-deviation increase in vacancy duration, measured in focal-team games. Outcomes are pooled OPS, total bases per plate appearance (TB/PA), and times on base per plate appearance (TOB/PA) for the fixed worker pool. The comparison is return games 0 through 20 relative to the clean pre-event period, event-time games $-5$ through $-2$; treated return follow-up ends if the original focal high performer becomes unavailable again or is subsequently observed playing for another team. Models use event-balanced unit weights, event-unit fixed effects, and event-by-period fixed effects. Standard errors, two-way clustered by focal event and team-season, are in parentheses.
\end{minipage}
\end{table}

\clearpage
\subsection{Alternative Performance Signals}
\label{app:alternative-signals}

\begin{table}[!htbp]
\centering
\caption{Alternative Measures of Vacancy-Period Performance Surprise}
\label{tab:appendix-signals}
\begin{adjustbox}{max width=\textwidth}
\begin{tabular}{lcccc}
\toprule
 & \shortstack{Post-return\\PA/game} & \shortstack{Career\\continuation} & \shortstack{Next-season\\PA} & \shortstack{Next-season\\starts} \\
 & (1) & (2) & (3) & (4) \\ \midrule
\multicolumn{5}{l}{\textit{Panel A. Recent OPS surprise (baseline)}} \\
Performance surprise & 0.160*** & 0.033*** & 14.746*** & 3.158*** \\
 & (0.029) & (0.008) & (3.788) & (0.868) \\
R-squared & 0.331 & 0.300 & 0.314 & 0.305 \\
Observations & 430 & 430 & 430 & 430 \\
\multicolumn{5}{l}{\textit{Panel B. Career OPS surprise}} \\
Performance surprise & 0.163*** & 0.034*** & 15.084*** & 3.229*** \\
 & (0.030) & (0.008) & (3.896) & (0.891) \\
R-squared & 0.312 & 0.293 & 0.297 & 0.288 \\
Observations & 430 & 430 & 430 & 430 \\
\multicolumn{5}{l}{\textit{Panel C. Recent TB/PA surprise}} \\
Performance surprise & 0.244*** & 0.057*** & 23.677*** & 5.167*** \\
 & (0.054) & (0.012) & (6.856) & (1.565) \\
R-squared & 0.308 & 0.296 & 0.302 & 0.295 \\
Observations & 430 & 430 & 430 & 430 \\
\multicolumn{5}{l}{\textit{Panel D. Recent TOB/PA surprise}} \\
Performance surprise & 0.361*** & 0.060*** & 28.790*** & 6.010*** \\
 & (0.063) & (0.019) & (8.632) & (1.969) \\
R-squared & 0.329 & 0.281 & 0.302 & 0.291 \\
Observations & 430 & 430 & 430 & 430 \\
\bottomrule
\end{tabular}

\end{adjustbox}

\medskip
\begin{minipage}{0.96\textwidth}
\footnotesize
\textit{Notes:} Panel A reproduces the recent-OPS specification. Panel B uses career-to-date OPS as the prior measure; Panels C and D use recent total bases per plate appearance (TB/PA) and times on base per plate appearance (TOB/PA), respectively. Each performance surprise is vacancy performance minus its corresponding pre-event measure. Coefficients are scaled to a +0.100 increase in the corresponding performance surprise. Outcomes, controls, event-season, focal-position, and treated-team fixed effects, and equal worker weighting within event follow Table~\ref{tab:performance-signals}. Standard errors, two-way clustered by focal event and player, are in parentheses. $^{*}p<0.10$, $^{**}p<0.05$, $^{***}p<0.01$.
\end{minipage}
\end{table}

\end{document}